\documentclass{appolb}
\usepackage{graphicx}
\usepackage{amsmath,amssymb}
\usepackage{array}
\usepackage{ragged2e}
\newcolumntype{P}[1]{>{\RaggedRight\hspace{0pt}}p{#1}}
\usepackage[
backend=biber,
style=numeric,
sorting=none
]{biblatex}
\renewbibmacro{in:}{\addcomma\space}
\DeclareFieldFormat{pages}{#1}
\DeclareDelimFormat[bib,biblist]{nameyeardelim}{\addcomma\space}
\AtEveryBibitem{\clearfield{title}}
\AtEveryBibitem{\clearfield{doi}}

\begin{document}
\title{Experimental overview of electromagnetic probes in ultra-relativistic nucleus-nucleus collisions%
\thanks{Presented at the Quark Matter 2022 conference in Krak\'{o}w}%
}
\author{Klaus Reygers
\address{Physikalisches Institut, Heidelberg University, Germany}
\\[3mm]
}
\maketitle
\begin{abstract}
Electromagnetic probes are not affected by hadronization and provide direct information about the space-time evolution of high-energy  nucleus-nucleus collisions. In particular, the measurement of thermal radiation from the quark-gluon plasma and the extraction of an effective medium temperature belong to the key objectives in heavy-ion physics. We provide a brief tour of current results and an outlook to future measurements. 
\end{abstract}
  
\section{Introduction}
Once electroweak probes ($\gamma$, $\gamma^*$, W, Z) are produced in a high-energy nucleus-nucleus collision, they leave the surrounding medium with virtually no further interactions. The direct information these probes provide about the medium is therefore different from the information from light and heavy quark observables which are affected by hadronization. Electroweak bosons provide information about the entire evolution of a nucleus-nucleus collision. Parton distribution functions in nuclei are probed by measuring $\gamma$, $\gamma^*$, W, Z produced in initial hard scattering processes. Moreover, parton energy loss in the medium can be studied by measuring quark jets recoiling from a photon or a Z boson. Real and virtual photons produced in the preequilibrium phase, the quark-gluon plasma (QGP), and the hadron gas provide information about the space-time evolution and the temperature of the medium. In addition, the broadening of the $\rho$ resonance in nucleus-nucleus collisions accessible via the decay $\rho \to e^+e^-$ is sensitive to the restoration of chiral symmetry \cite{Rapp:2010sj}. This paper focuses on thermal radiation from the hot medium created in high-energy nucleus-nucleus collision.

Thermal photons from the QGP are expected to be a significant component of the direct-photon yield in the transverse momentum range $1 \lesssim p_T \lesssim 3~\mathrm{GeV}/c$ \cite{David:2019wpt,Monnai:2022hfs}. The inverse slope parameter $T_\mathrm{eff}$ is sensitive to the effective medium temperature averaged over the space-time evolution of the collision. Due to the rapid radial expansion of the medium, the measured photon spectrum is blueshifted and the inverse slope parameter is expected to be larger than the temperature of the emitting medium. Therefore, an inverse slope parameter above the critical temperature $T_\mathrm{pc} \approx 156~\mathrm{MeV}$ cannot be directly interpreted as evidence for the formation of a QGP. With increasing $p_T$, the measured direct-photon yield becomes more sensitive to earlier and hotter stages of the collisions \cite{Shen:2013vja}. Photons from the preequilibrium stage could become important for $p_T \gtrsim 2-3~\mathrm{GeV}/c$ \cite{Gale:2021emg}.

Dileptons (virtual photons) probe different aspects of a heavy-ion collisions depending on the considered mass range. Dileptons with invariant masses around and below the $\rho$ meson mass are particularly sensitive to modifications of the $\rho$ meson due to the surrounding medium. These modifications are experimentally accessible as the $\rho$ decays in the medium due to its short lifetime of only $1.3~\mathrm{fm}/c$. In particular the presence of baryons in the medium is expected to give rise to $\rho$ melting, i.e., to a strong increase of its width. The $\rho$ melting is connected to the restoration of chiral symmetry in a hot and dense medium. The intermediate mass region (IMR, $1 \lesssim m_{ee} \lesssim 3\,\mathrm{GeV}/c^2$) is sensitive to thermal radiation from the QGP. An effective medium temperature not affected by blueshift can be obtained by fitting the dielectron yield in this range with $dN/m_{ee} \propto (m_{ee}T)^{3/2} \exp(-m_{ee}/T)$. With increasing mass, the contribution from dielectrons from the preequilibrium stage becomes more relevant which needs to be considered in the extraction of a temperature. To date, the only measurement of a medium temperature in the mass range above $1\,\mathrm{GeV}/c^2$ was made by the NA60 experiment at the CERN SPS in In-In collisions collisions. A fit in the range $1.2 \lesssim m_{ee} \lesssim 2.0\,\mathrm{GeV}/c^2$ gave an effective temperature of $T = 205 \pm 12\,\mathrm{MeV}$ \cite{Specht:2010xu}. 

\section{Low-$p_T$ direct photons}
Direct photon measurements are expected to eventually play a key role in our understanding of the evolution of the QGP in nucleus-nucleus collisions. However, it is currently a challenge for models to simultaneously describe the direct-photons yield and the azimuthal anisotropy ($v_2$) of direct photons. This is known as the direct-photon puzzle \cite{Paquet:2015lta,Gale:2021emg}. The puzzle became apparent at the Quark Matter 2011 conference with the observation by the PHENIX collaboration that the $v_2$ of direct photons was similar to the $v_2$ of pions whereas hydrodynamic models predicted a much smaller direct-photon $v_2$. In addition, hydrodynamic models underestimated the direct photon yield at low $p_T$ by about a factor 2--3 \cite{Paquet:2015lta}. The radial flow velocity profiles created in the hydrodynamic evolution of the medium need time to build up. The large direct-photon $v_2$ therefore could mean that contrary to the general expectation, direct photons in the range $1 \lesssim p_T \lesssim 3~\mathrm{GeV}/c$ are predominantly produced at the later stages of the medium evolution, perhaps during the transition from the QGP to a gas of hadrons. This would make direct photons a less useful probe of the early hot QGP phase. The direct-photon puzzle currently is a puzzle which concerns the PHENIX spectra and $v_2$ measurements. At the LHC, ALICE direct-photon spectra and $v_2$ measurements show similar trends as the PHENIX data, however, the uncertainties are unfortunately too large to claim a significant difference between data and theory.

\begin{figure}[t]
\centerline{%
\includegraphics[width=0.45\textwidth]{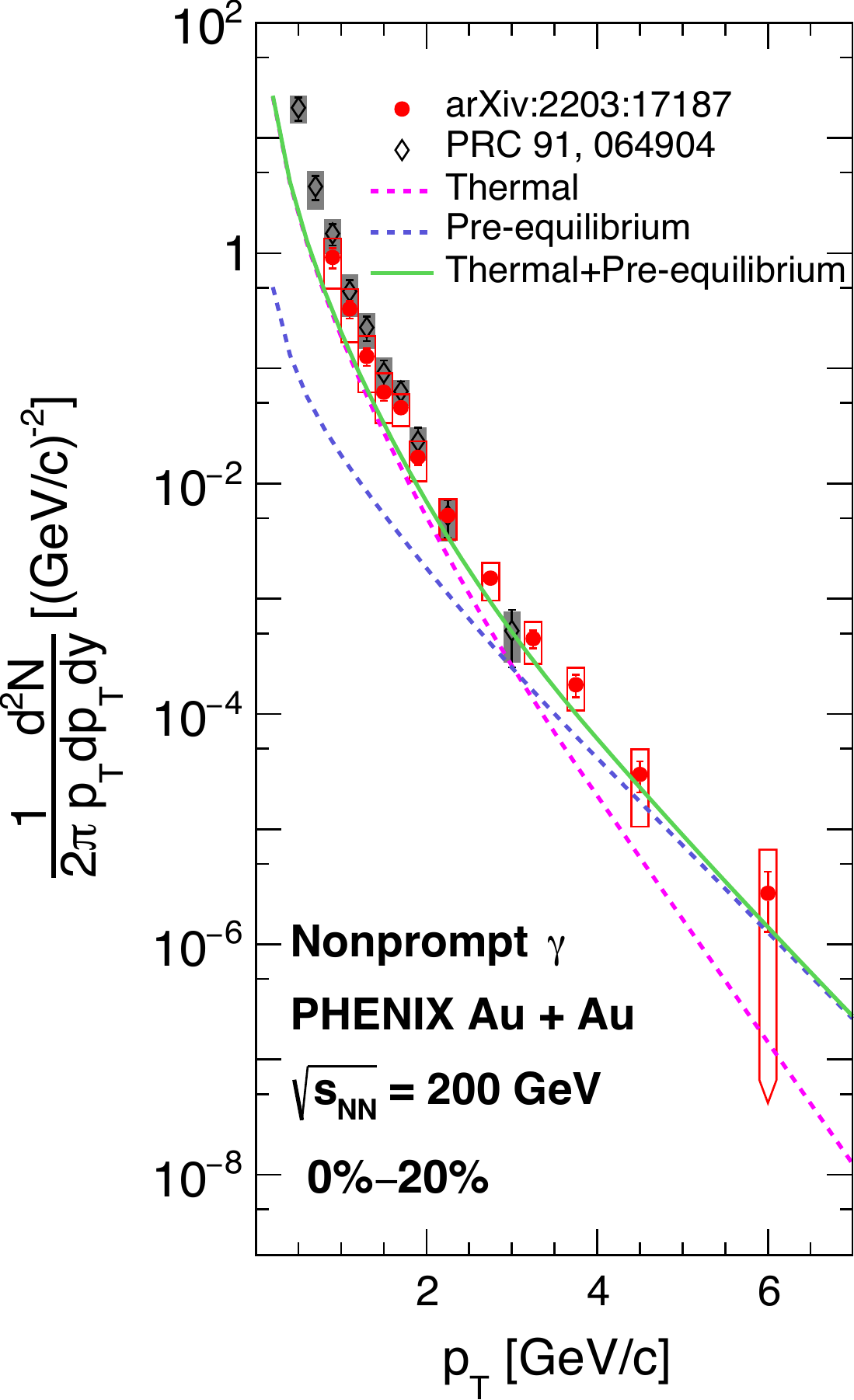}
\hfill
\raisebox{0.15\textwidth}{\includegraphics[width=0.5\textwidth]{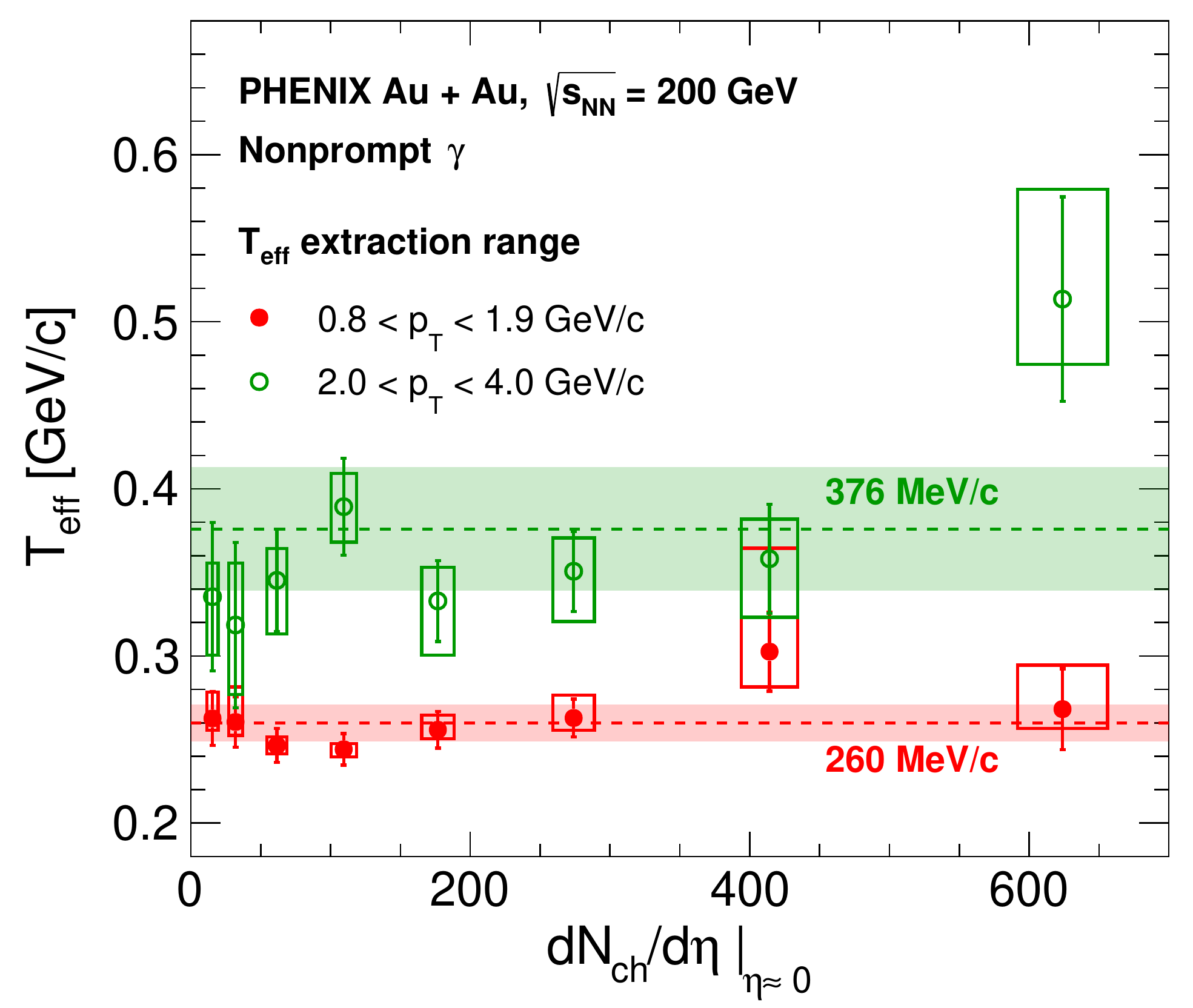}}}
\caption{Left panel: Nonprompt (``$\text{Au--Au} - N_\mathrm{coll} \times \text{pp}$'') direct-photon spectrum in central Au--Au collisions at $\sqrt{s_\mathrm{NN}} = 200~\mathrm{GeV}$ \cite{PHENIX:2022rsx}. The new data from the high-statistics 2014 Au+Au run are shown as closed circles. Right panel: Inverse slope parameters $T_\mathrm{eff}$ in Au--Au collisions at $\sqrt{s_\mathrm{NN}} = 200~\mathrm{GeV}$ for different centralities characterized by the charged-particle multiplicity density $\left. dN_\mathrm{ch}/d\eta \right|_{\eta \approx 0}$.}
\label{fig:phenix_photon_spectrum_and_Teff}
\end{figure}
At this conference, PHENIX presented final direct-photon spectra in Au--Au collisions at $\sqrt{s_\mathrm{NN}} = 200\,\mathrm{GeV}$ from the high-statistics 2014 data set \cite{PHENIX:2022rsx}. A significant excess above the direct-photon spectrum in pp collisions at $\sqrt{s} = 200\,\mathrm{GeV}$ scaled by the number of nucleon-nucleon collisions ($N_\mathrm{coll}$) is found. The difference between in direct-photon spectra in Au--Au collisions and the scaled pp spectrum, denoted by PHENIX as ``nonprompt'' direct-photon spectrum, is shown in Fig.~\ref{fig:phenix_photon_spectrum_and_Teff}. PHENIX extracted the inverse slope parameter $T_\mathrm{eff}$ in two fit ranges for various centralities (see Fig.~\ref{fig:phenix_photon_spectrum_and_Teff}, right panel). The inverse slope parameter is independent of the centrality but increases with increasing $p_T$, in line with a larger early-time contribution at higher $p_T$.

\begin{figure}[t]
\centerline{%
\includegraphics[width=0.535\textwidth]{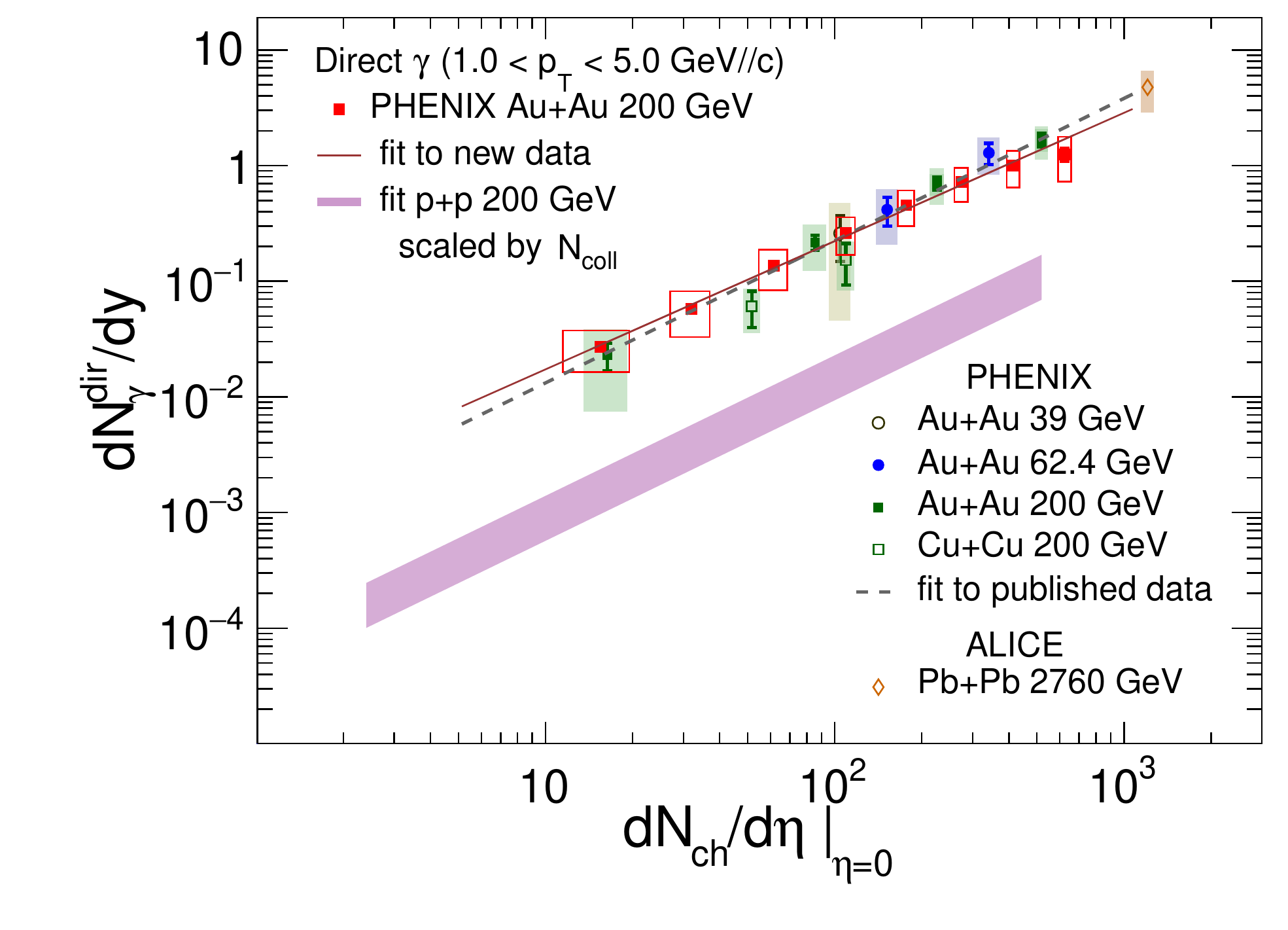}
\hfill
\raisebox{0.0072\textwidth}{\includegraphics[width=0.468\textwidth]{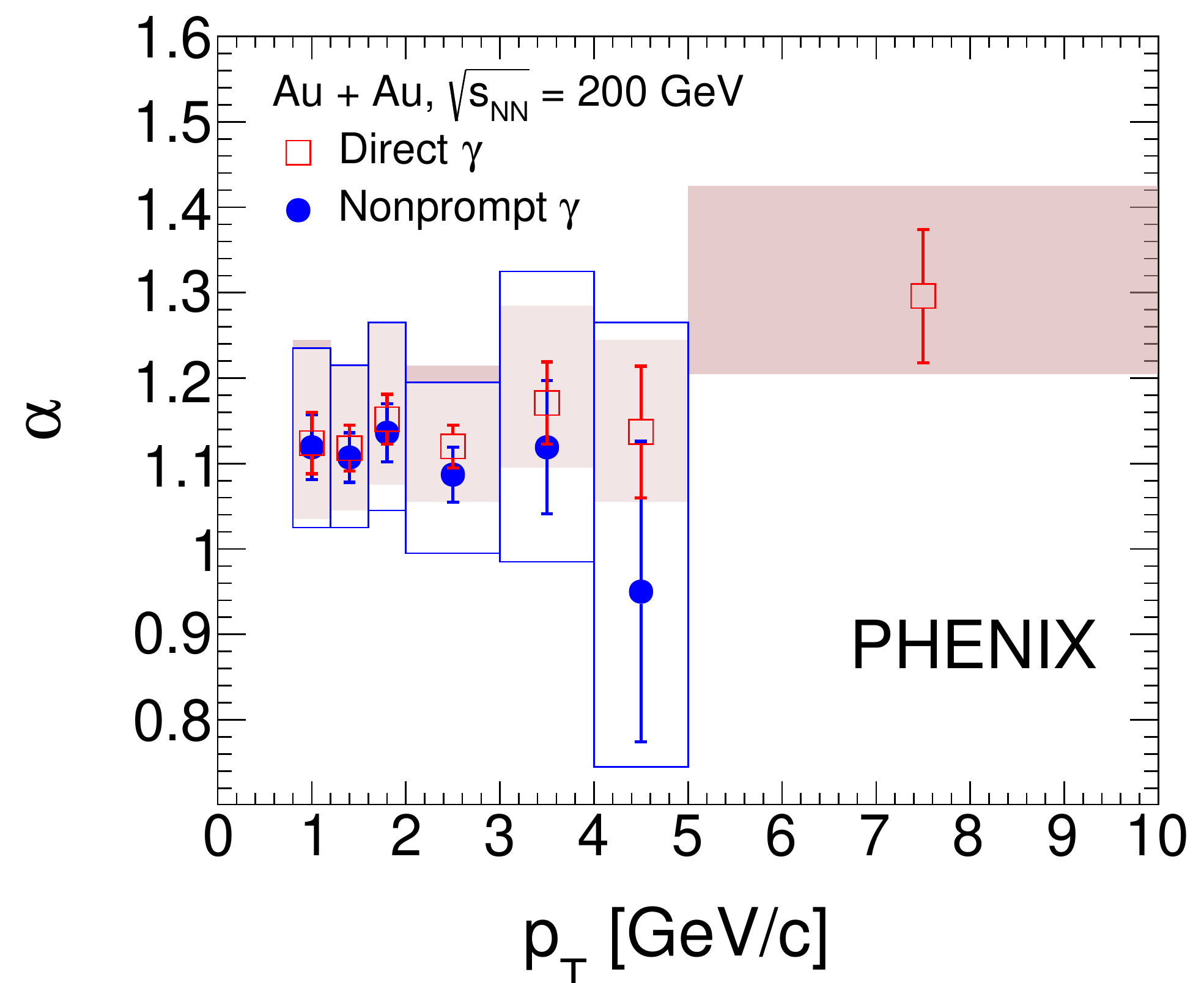}}}
\caption{Left panel: Direct photon yield $dN_\gamma^\mathrm{dir}/dy$ in the range $1 < p_T < 5\,\mathrm{GeV}/c$ for collisions with different charged-particle multiplicity densities $\left. dN_\mathrm{ch}/d\eta \right|_{\eta \approx 0}$. Right panel: Scaling power $\alpha$ describing the charged-particle multiplicity dependence of the direct-photon yield in different $p_T$ intervals according to $dN_\gamma^\mathrm{dir}/dy \propto (\left. dN_\mathrm{ch}/d\eta \right|_{\eta \approx 0})^\alpha$.}
\label{fig:phenix_scaling}
\end{figure}

To examine at which stage of a nucleus-nucleus collision direct photons are predominantly produced, it is instructive to study the dependence of the direct-photon yield on centrality and center-of-mass energy. PHENIX found a universal scaling according to which the direct-photon yield integrated above $p_T = 1\,\mathrm{GeV}/c$ is a function of the charged-particle multiplicity alone (for different centralities, center-of-mass energies, and collision systems) and is given by $dN_\gamma^\mathrm{dir}/dy \propto (\left. dN_\mathrm{ch}/d\eta \right|_{\eta \approx 0})^\alpha$, see Fig.~\ref{fig:phenix_scaling}. A fit to the 2014 data gives $\alpha = 1.11 \pm 0.02\,(\mathrm{stat}) {}^{+0.09}_{-0.08}\,(\mathrm{syst})$. Interestingly, this value is smaller than the power $\alpha \approx 1.6$ expected for thermal photons in \cite{Shen:2013vja}. Photons from different stages of a collision are actually expected to exhibit a different scaling behavior. In \cite{Shen:2013vja} it is predicted that the hadron gas, the QGP, and hard scattering contributions scale as $\alpha_\mathrm{HG} \approx 1.23$, $\alpha_\mathrm{QGP} \approx 1.83$, and $\alpha_\mathrm{pQCD} \approx 1.25$, respectively. Different stages contribute differently to different $p_T$ intervals and one might expect a strong $p_T$ dependence of $\alpha$. PHENIX, however, finds that $\alpha(p_T)$ is consistent with being independent of $p_T$.

\begin{figure}[t]
\centerline{%
\includegraphics[width=.8\textwidth]{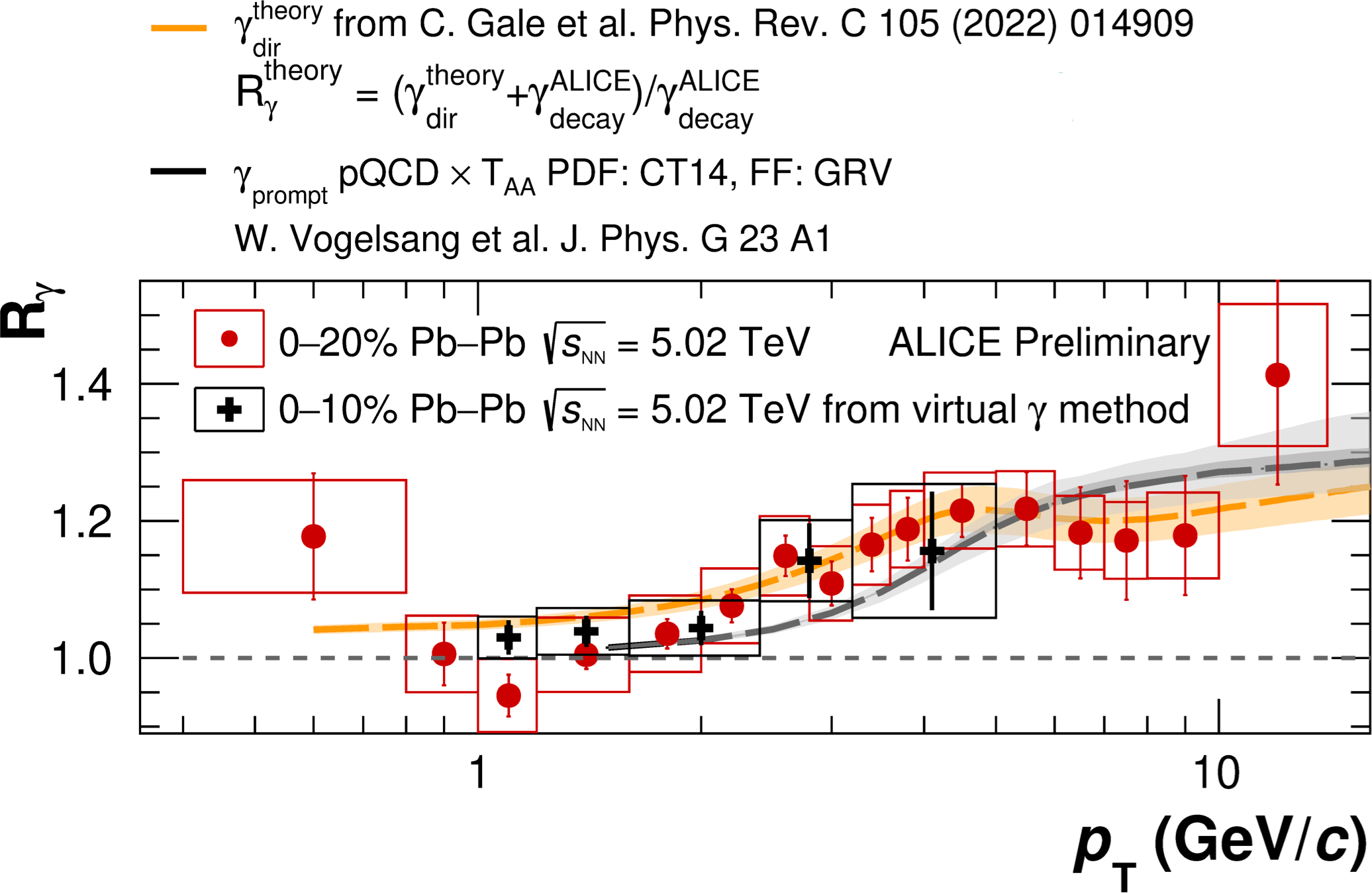}}
\caption{Direct-photon excess ratio $R_\gamma = \gamma_\mathrm{total} / \gamma_\mathrm{decay} \equiv 1 + \gamma_\mathrm{direct}/\gamma_\mathrm{decay}$ in central Pb--Pb collisions at $\sqrt{s_\mathrm{NN}} = 5.02~\mathrm{TeV}$. A scaled pQCD calculation (gray dashed line) and a model with additional photon production sources including thermal radiation fom the QGP (orange dashed line) are compared to the data.}
\label{fig:ALICE_Rgamma}
\end{figure}

ALICE showed new preliminary direct-photon spectra in Pb--Pb collisions at $\sqrt{s_\mathrm{NN}} = 5.02\,\mathrm{TeV}$ \cite{Danisch}. The results of two independent methods, the photon conversion method and the virtual photons method, were found to be in agreement, see Fig.~\ref{fig:ALICE_Rgamma}. Photons from initial hard parton-parton scatterings as obtained from a scaled next-to-leading-order perturbative QCD (pQCD) calculation are sufficient to explain the data. A calculation with additional contributions from thermal and preequilibrium photons also agrees with the data. To establish a signal of thermal photons in nucleus-nucleus collisions at the LHC thus requires a further reduction of the experimental uncertainties.

\begin{figure}[t]
\centerline{%
\includegraphics[width=.7\textwidth]{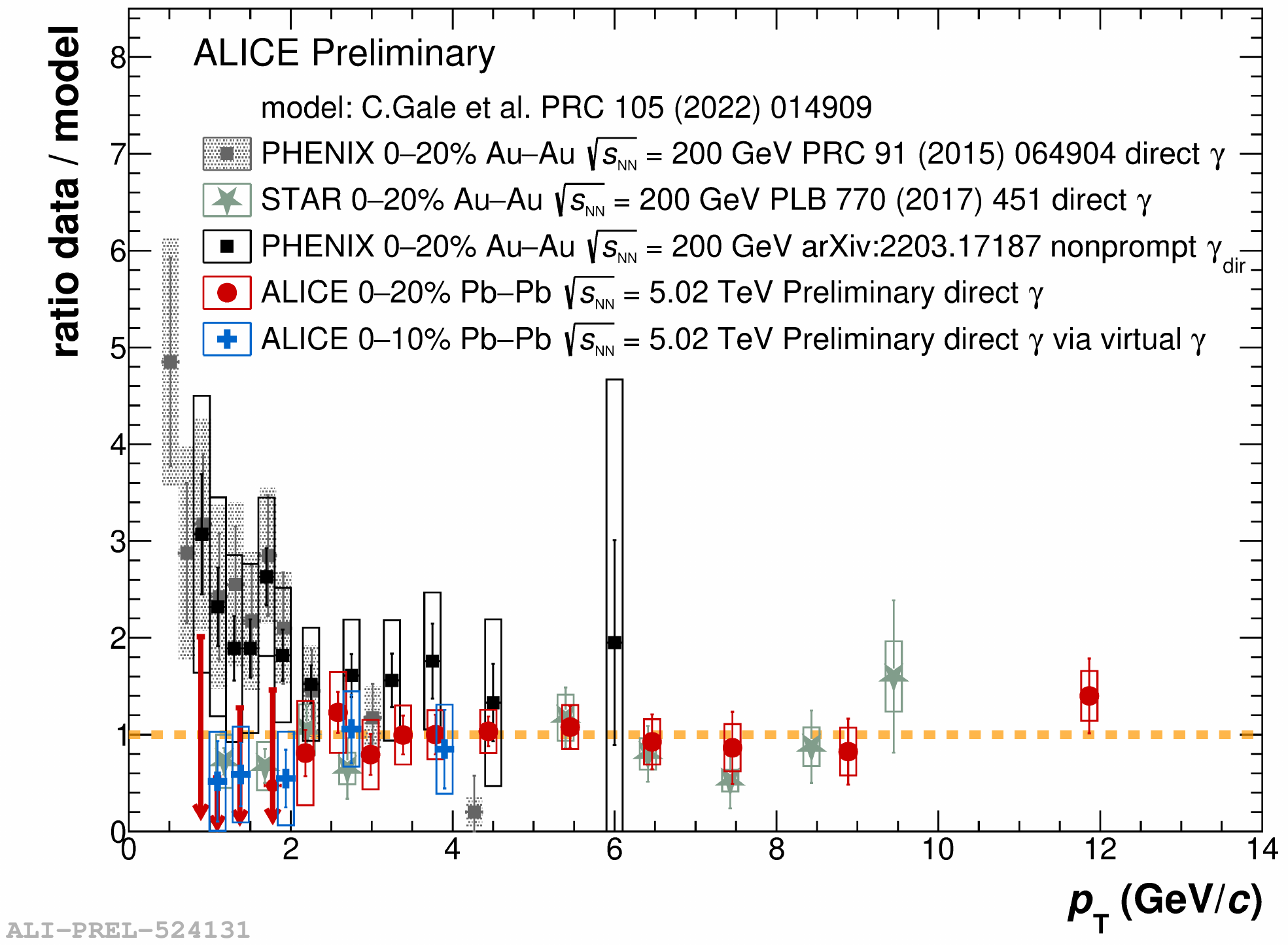}}
\caption{Comparison of direct-photon transverse momentum spectra measured at RHIC and the LHC to a model which includes thermal and preequilibrium photons in addition to pQCD photons.}
\label{fig:direct_photon_yield_over_model}
\end{figure}

With the calculation of \cite{Gale:2021emg} and the new direct-photon spectra presented at this conference, the differences between data and theory in terms of the yields do not appear very significant anymore, see Fig.~\ref{fig:direct_photon_yield_over_model}. At low $p_T \lesssim 3~\mathrm{GeV}/c$ where thermal and preequilibrium photons are expected, one finds a good agreement of the model calculation with the LHC measurements. At RHIC, the model predictions agrees with the STAR data \cite{STAR:2016use} at low $p_T$. The calculation is systematically below the PHENIX data, however, within the experimental systematic uncertainties there is no significant deviation. While the puzzle appears almost resolved concerning the yields, it remains a challenge to explain the large azimuthal anisotropy ($v_2$) measured by PHENIX.

\section{Dileptons}
\begin{figure}[t]
\centerline{%
\includegraphics[width=.7\textwidth]{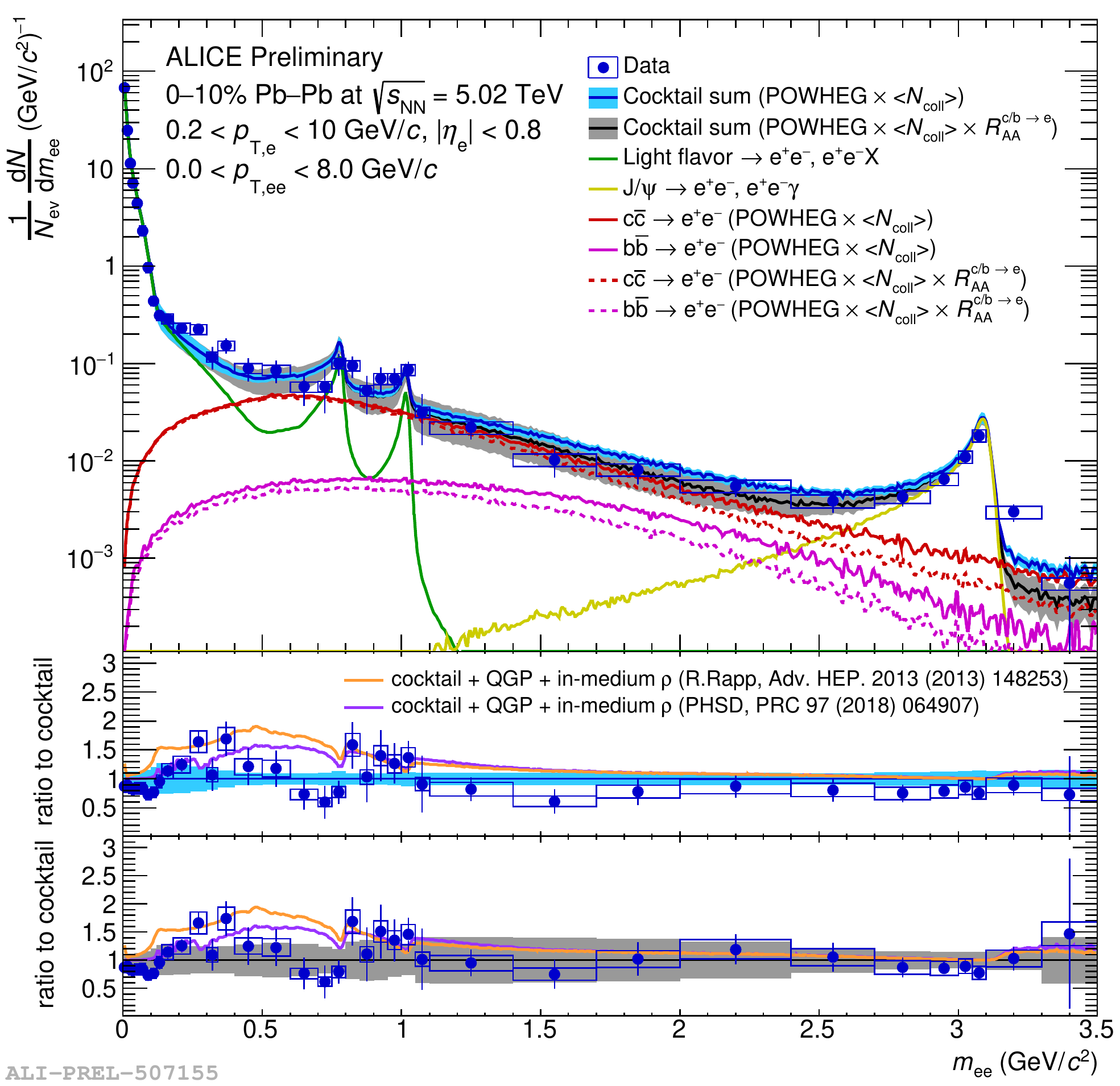}}
\caption{Dielectron invariant mass distribution in central Pb--Pb collisions at $\sqrt{s_\mathrm{NN}} = 5.02~\mathrm{TeV}$ after subtraction of uncorrelated background. The data are compared to the expected yield (``cocktail'') from the decay of hadrons (light flavor hadrons and $J/\psi$) and from $e^+e^-$ pairs resulting from correlated semileptonic open charm and open beauty hadron decays. The two lower panels show the comparison to two versions of the cocktail, with and without nuclear modifications of the $c \bar{c} \to e^+e^-$ and $b \bar{b} \to e^+e^-$ contribution.}
\label{fig:ALICE_dielectron_minv}
\end{figure}

ALICE presented a preliminary dielectron mass spectrum in central Pb--Pb collisions at $\sqrt{s_\mathrm{NN}} = 5.02~\mathrm{TeV}$ at this conference, see Fig.~\ref{fig:ALICE_dielectron_minv} \cite{Jung}. To establish a signal of thermal radiation in the low mass region (LMR, $m_{ee} \lesssim 1\,\mathrm{GeV}/c^2$) and in the intermediate mass region (IMR, $1 \lesssim m_{ee} \lesssim 3\,\mathrm{GeV}/c^2$) at LHC energies, a detailed understanding of $e^+e^-$ pairs resulting from correlated semileptonic decays of open charm and beauty hadrons is crucial ($c \bar{c} \to e^+e^-$, $b \bar{b} \to e^+e^-$). In Fig.~\ref{fig:ALICE_dielectron_minv} two versions of the cocktail, with and without nuclear modification of the $c \bar{c} \to e^+e^-$ and $b \bar{b} \to e^+e^-$ contribution are compared to the data. In both cases no clear enhancement above the cocktail is observed. However, predictions that include thermal QGP radiation and a medium-modified $\rho$ also appear to be consistent with the data. Separation of the prompt thermal signal from the non-prompt background of semileptonic decays based on the distance of closest approach of the electron and positron tracks to the primary interaction point will play a crucial role in the extraction of a thermal signal.

\section{Future measurements}
One of the main current goals in the study of electromagnetic probes is to obtain a global picture of high-energy nucleus-nucleus collisions in which both electromagnetic observables as well as light and heavy quark observables can be naturally explained. This entails the determination of a unique signal of thermal radiation as well as the extraction of an effective medium temperature. An effective medium temperature above the critical temperature $T_\mathrm{pc} \approx 156\,\mathrm{MeV}$ would provide additional evidence for the formation of a QGP. To improve our understand of the QCD phase diagram, future measurements of electromagnetic probes at vanish and high baryo-chemical potential $\mu_\mathrm{B}$ play a crucial role, see Tab.~\ref{tab:future_measurments}. These include, e.g., the study of the caloric curve ($T_\mathrm{eff}$ as a fct.\ of $\sqrt{s_\mathrm{NN}}$) to find evidence for a first-order phase transition at high $\mu_B$. In addition, real and virtual photons at very low $p_T$ (and masses) could provide information about the electrical conductivity of the QGP. Moreover, the long-standing puzzle of the measured excess of ultra-soft photons above predictions based on Low's theorem \cite{Low:1958sn} should finally be resolved. 
\begin{table}[t]
\begin{center}
\begin{tabular}{P{0.34\linewidth} P{0.27\linewidth} P{0.27\linewidth}}
 & High $\mu_\mathrm{B}$ & Vanishing $\mu_\mathrm{B}$ \\
 \hline 
 Thermal radiation \newline ($\gamma$, $\gamma^*$ spectra) & STAR, NA60+, CBM, NICA & PHENIX, STAR, \mbox{ALICE~2}, ALICE~3 \\
 \hline  
 Caloric curve \newline ($T_\mathrm{eff}$ in IMR vs.\ $\sqrt{s_\mathrm{NN}}$) & STAR, NA60+, CBM, NICA & \\
 \hline
 Mechanism of chiral symmetry resoration, $\rho$-$a_1$ mixing & CBM, NA60+ & STAR, ALICE~3 \\
 \hline
 Dilepton + direct photon $v_n$ & CBM, HADES, NA60+ & STAR, (ALICE~2), ALICE~3 \\ 
 \hline
 Time dep.\ of $T_\mathrm{eff}$, \newline pre\-equilibrium contrib.\ & & ALICE~3\\
 \hline
 Ultra-soft photons \newline (test of Low’s theorem) & & ALICE~3 \\
 \hline
 Electrical conductivity & HADES & STAR, ALICE~3 \\ 
\end{tabular}
\end{center}
\caption{Possible future measurements of electromagnetic observables at high and vanishing baryochemical potential $\mu_\mathrm{B}$ along with the corresponding experiments \cite{na60plus,ALICE3}.}
\label{tab:future_measurments}
\end{table}

\printbibliography

@article{Danisch,
    author = "Danisch, Meike",
    collaboration = "ALICE",
    eprint = "these_proceedings",
    month = "8",
    year = "2022"
}

@article{Jung,
    author = "Jung, Jerome",
    collaboration = "ALICE",
    eprint = "these_proceedings",
    month = "8",
    year = "2022"
}

@article{PHENIX:2022rsx,
    author = "PHENIX collaboration",
    collaboration = "PHENIX",
    title = "{Nonprompt direct-photon production in Au$+$Au collisions at $\sqrt{s_{_{NN}}}=200$ GeV}",
    eprint = "2203.17187",
    archivePrefix = "arXiv",
    primaryClass = "nucl-ex",
    month = "3",
    year = "2022"
}

@article{Paquet:2015lta,
    author = {Paquet, Jean-Fran\c{c}ois and Shen, Chun and Denicol, Gabriel S. and Luzum, Matthew and Schenke, Bj\"orn and Jeon, Sangyong and Gale, Charles},
    title = "{Production of photons in relativistic heavy-ion collisions}",
    archivePrefix = "arXiv",
    primaryClass = "hep-ph",
    doi = "10.1103/PhysRevC.93.044906",
    journal = "Phys. Rev. C",
    volume = "93",
    number = "4",
    pages = "044906",
    year = "2016"
}

@article{Gale:2021emg,
    author = {Gale, Charles and Paquet, Jean-Fran\c{c}ois and Schenke, Bj\"orn and Shen, Chun},
    title = "{Multimessenger heavy-ion collision physics}",
    archivePrefix = "arXiv",
    primaryClass = "nucl-th",
    doi = "10.1103/PhysRevC.105.014909",
    journal = "Phys. Rev. C",
    volume = "105",
    number = "1",
    pages = "014909",
    year = "2022"
}

@article{David:2019wpt,
    author = "David, Gabor",
    title = "{Direct real photons in relativistic heavy ion collisions}",
    archivePrefix = "arXiv",
    primaryClass = "nucl-ex",
    doi = "10.1088/1361-6633/ab6f57",
    journal = "Rept. Prog. Phys.",
    volume = "83",
    number = "4",
    pages = "046301",
    year = "2020"
}

@article{STAR:2016use,
    author = "STAR collaboration",
    collaboration = "STAR",
    title = "{Direct virtual photon production in Au+Au collisions at $\sqrt{s_{NN}}$ = 200 GeV}",
    archivePrefix = "arXiv",
    primaryClass = "nucl-ex",
    doi = "10.1016/j.physletb.2017.04.050",
    journal = "Phys. Lett. B",
    volume = "770",
    pages = "451--458",
    year = "2017"
}

@article{Monnai:2022hfs,
    author = "Monnai, Akihiko",
    title = "{Direct photons in hydrodynamic modeling of relativistic nuclear collisions}",
    archivePrefix = "arXiv",
    primaryClass = "nucl-th",
    doi = "10.1142/S0217751X2230006X",
    journal = "Int. J. Mod. Phys. A",
    volume = "37",
    number = "11n12",
    pages = "2230006",
    year = "2022"
}

@article{Specht:2010xu,
    author = "Specht, Hans J.",
    collaboration = "NA60",
    title = "{Thermal Dileptons from Hot and Dense Strongly Interacting Matter}",
    archivePrefix = "arXiv",
    primaryClass = "nucl-ex",
    doi = "10.1063/1.3541982",
    journal = "AIP Conf. Proc.",
    volume = "1322",
    number = "1",
    pages = "1--10",
    year = "2010"
}

@article{Shen:2013vja,
    author = "Shen, Chun and Heinz, Ulrich W and Paquet, Jean-Francois and Gale, Charles",
    title = "{Thermal photons as a quark-gluon plasma thermometer reexamined}",
    archivePrefix = "arXiv",
    primaryClass = "nucl-th",
    doi = "10.1103/PhysRevC.89.044910",
    journal = "Phys. Rev. C",
    volume = "89",
    number = "4",
    pages = "044910",
    year = "2014"
}

@techreport{na60plus,
    author = "NA60+ collaboration",
    title = "{Expression of Interest for a new experiment at the CERN SPS: NA60+}",
    number      = {CERN-SPSC-2019-017; SPSC-EOI-019},
    institution = {CERN},
    year        = {2019}
}

@techreport{ALICE3,
    author = "ALICE collaboration",
    title = "{Letter of intent for ALICE 3: A next generation heavy-ion experiment at the LHC}",
    number      = {CERN-LHCC-2022-009; LHCC-I-038},
    institution = {CERN},
    year        = {2022}
}

@article{Rapp:2010sj,
    author = "Rapp, Ralf",
    title = "{In-Medium Vector Mesons, Dileptons and Chiral Restoration}",
    archivePrefix = "arXiv",
    primaryClass = "nucl-th",
    doi = "10.1063/1.3542032",
    journal = "AIP Conf. Proc.",
    volume = "1322",
    number = "1",
    pages = "55--63",
    year = "2010"
}

@article{Low:1958sn,
    author = "Low, F. E.",
    title = "{Bremsstrahlung of very low-energy quanta in elementary particle collisions}",
    doi = "10.1103/PhysRev.110.974",
    journal = "Phys. Rev.",
    volume = "110",
    pages = "974--977",
    year = "1958"
}

\end{document}